\documentclass[preprint,superscriptaddress,showpacs,nofootinbib,aps]{revtex4}
 \usepackage{graphicx}
 \usepackage{bm}
 
\begin{document}

 \title{Study of $B_{c}$ ${\to}$ $B^{({\ast})}P$, $BV$ Decays with QCD Factorization}
 \author{Junfeng Sun}
 \affiliation{College of Physics and Information Engineering,
              Henan Normal University,
              Xinxiang 453007, China}
 \thanks{Mailing address}
 \affiliation{Theoretical Physics Center for Science Facilities (TPCSF),
 Institute of High Energy Physics,
 Chinese Academy of Sciences (IHEP, CAS)}

 \author{Yueling Yang}
 \affiliation{College of Physics and Information Engineering,
              Henan Normal University,
              Xinxiang 453007, China}

 \author{Wenjie Du}
 \affiliation{College of Physics and Information Engineering,
              Henan Normal University,
              Xinxiang 453007, China}

 \author{Huilan Ma}
 \affiliation{College of Physics and Information Engineering,
              Henan Normal University,
              Xinxiang 453007, China}

 \begin{abstract}
 The $B_{c}$ ${\to}$ $B^{({\ast})}_{q}P$, $B_{q}V$ decays are
 studied with the QCD factorization approach (where $P$ and
 $V$ denote pseudoscalar and vector mesons, respectively; $q$ $=$ $u$,
 $d$ and $s$). Considering the contributions of both current-current
 and penguin operators, the amplitudes of branching ratios are estimated
 at the leading approximation. We find that the contributions of the
 penguin operators are very small due to the serious suppression by
 the CKM elements. The most promising decay modes are $B_{c}$ ${\to}$
 $B_{s}^{({\ast})}{\pi}$, $B_{s}{\rho}$, which might be easily
 detected at hadron colliders.
 \end{abstract}

 \maketitle

 \section{Introduction}
 \label{sec1}
 The $B_{c}$ meson is one of the unique ``double heavy-flavored'' binding
 system in the standard model (SM). The study of the $B_{c}$ meson has
 received a great interest, due to its special properties:
 (1) The $B_{c}$ meson carries open flavors. We can study the two heavy flavors
 of both $b$ and $c$ quarks simultaneously with the $B_{c}$ meson.
 (2) The $B_{c}$ meson can serve as a great laboratory for potential models,
 QCD sum rules, Heavy Quark Effective Theory (HQET), lattice QCD, etc.
 (3) The $B_{c}$ meson has rich decay channels, because of its sufficiently
 large mass and that the $b$ and $c$ quarks can decay individually. The
 $B_{c}$ meson decays may provide windows for testing the predictions of the
 SM and can shed light on new physics beyond SM.

 The $B_{c}$ mesons are too massive to access at the $B$-factories near
 ${\Upsilon}(4S)$. They can be produced in significant numbers at hadron
 colliders. The $B_{c}$ meson has been firstly discovered by the CDF
 Collaboration \cite{cdf98}. Recently the CDF and D0 Collaborations announced
 some accurate measurements \cite{cdf07,d008} with part of their available
 data. Much more $B_{c}$ mesons and detailed information about their
 decay properties are expected at the Large Hadron Collider (LHC) which
 is scheduled to run in this year. It is estimated that one could expect
 around $5$ ${\times}$ $10^{10}$ $B_{c}$ events per year at LHC
 \cite{0412158,pan67p1559} due to the relatively large production cross
 section \cite{prd71p074012} plus the huge luminosity ${\cal L}$ $=$
 $10^{34}\,{\rm cm}^{-2}{\rm  s}^{-1}$ and high center-of-mass energy
 $\sqrt{s}$ $=$ $14$ TeV \cite{lhc}. There seems to exist
 a real possibility to study not only some $B_{c}$ rare decays, but also
 $CP$ violation and polarization asymmetries. The study of the $B_{c}$
 meson will highlight the advantages of $B$ physics at hadron colliders.

 The $B_{c}$ meson is stable for strong interaction because it lies below
 the threshold of the $B$-$D$ mesons. The electromagnetic interaction cannot
 transform the $B_{c}$ meson into other hadrons containing both $b$ and $c$
 heavy quarks, because the $B_{c}$ meson itself is the ground state. The
 $B_{c}$ meson decays via weak interaction only, which can be divided into
 three classes:
 (1) the $b$ quark decay ($b$ ${\to}$ $c$, $u$) with $c$ quark as a spectator,
 (2) the $c$ quark decay ($c$ ${\to}$ $s$, $d$) with $b$ quark as a spectator,
 and (3) the weak annihilation channels. In the $B_{c}$ meson, both heavy quark
 can decay weakly, resulting in its much shorter lifetime than other $b$-flavored
 mesons, i.e. ${\tau}_{B_{c}}$ ${\lesssim}$ $\frac{1}{3}{\tau}_{B_{q}}$ (where
 $q$ $=$ $u$, $d$, and $s$) \cite{pdg2006}. Rates of the Class (1) and (2) are
 competitive in magnitude. The Cabibbo-Kobayashi-Maskawa (CKM) \cite{ckm}
 matrix elements ${\vert}V_{cb}{\vert}$ ${\ll}$ ${\vert}V_{cs}{\vert}$, that
 is in favor of the $c$-quark decay greatly, whereas the phase space factor
 $m_{b}^{5}$ ${\gg}$ $m_{c}^{5}$ compensates the CKM matrix elements a lot for
 the two flavors \cite{ijmpa21p777}. In fact, the dominant contributions to
 the $B_{c}$ lifetime comes from the $c$-quark decays [Class (2)] (${\approx}$ $70\%$),
 while the $b$-quark decay [Class (1)] and weak annihilation [Class (3)] are
 expected to add about $20\%$ and $10\%$, respectively \cite{0412158}.

 The $B_{c}$ meson decays have been widely studied in the literature due to
 some of its outstanding features. (1) The pure leptonic $B_{c}$ decays
 belong to the Class (3), which are free from strong interaction in final states
 and can be used to measure the decay constant $f_{B_{c}}$ and the CKM elements
 ${\vert}V_{cb}{\vert}$, but they are not fully reconstructed due to the
 missing neutrino. (2) The semileptonic $B_{c}$ decays provide an excellent
 laboratory to measure the CKM elements ${\vert}V_{cb}{\vert}$,
 ${\vert}V_{ub}{\vert}$, ${\vert}V_{cs}{\vert}$, ${\vert}V_{cd}{\vert}$ and
 form factors for transitions of $B_{c}$ ${\to}$ $b$- and $c$-flavored mesons.
 The first signal of $B_{c}$ is observed via this mode \cite{cdf98}. The most
 difficult theoretical work at present is how to evaluate the hadronic matrix
 elements properly and accurately. (3) The nonleptonic $B_{c}$ decays are
 the most complicated due to the participation of the strong interaction, which
 complicate the extraction of parameters in SM, but they also provide great
 opportunities to study perturbative and non-perturbative QCD, final state
 interactions, etc.

 The earlier nonleptonic decays of $B_{c}$ meson has been studied in
 \cite{0412158,pan67p1559,07070919,prd75p097304,prd73p054024,epjc45p711,
 prd70p074022,prd68p094020,epjc32p29,jpg28p595,jpg28p2241,prd65p114007,
 pan64p1860,pan64p2027,prd61p034012,prd62p057503,prd62p014019,cpl18p498,
 epjc9p557,epjc5p705,prd56p4133,plb387p187,pan60p1729,9605451,9504319,
 prd49p3399,prd46p3836,plb286p160,prd39p1342}.
 While $c$-quark decays take the lion's share of the $B_{c}$ lifetime,
 the study on the Class (2) has not received enough attention. This can be
 explained by the fact that on the one hand the available data on the $B_{c}$
 meson is very few, on the other hand it is assumed that the long distance
 effects and final state interferences might be quite huge and that the
 Class (2) decays were hard to detect experimentally.
 Accompanied by the LHC being about to run, the future copious data require
 more accurate theoretical predictions from now on.
 In this paper, we shall concentrate on the $B_{c}$ ${\to}$ $B_{q}^{({\ast})}P$,
 $B_{q}V$ (here $P$ and $V$ denote pseudoscalar and vector mesons, respectively;
 $q$ $=$ $u$, $d$ and $s$) decays in Class (2) with QCD factorization approach.
 Now let us outline a few reasons and arguments below.
 \begin{enumerate}
 \item From the experimental view
 \begin{itemize}
 \item The initial and final $b$-flavored mesons, i.e. the $B_{c}$ and
       $B_{q}^{({\ast})}$, all have a long lifetime due to their decays via the
       weak interaction. Considered the relativistic boost kinematically due to
       their large momentum obtained from huge center-of-mass energy, their
       information would be easily recorded by the multipurpose detectors sitting
       at the hadron colliders interaction regions (see \cite{lhc} for details).
 \item Although it is perceived that the hadron collider environment is ``messy''
       with high backgrounds, the $B_{c}$ ${\to}$ $B_{q}^{({\ast})}P$, $B_{q}V$
       decays are measurable due to the ``clean'' final states. Since the $B_{c}$
       meson carries charge, the final $B_{q}^{({\ast})}$ meson is tagged
       explicitly by the initial $B_{c}$ meson. The other light meson in the final
       state could also be identified effectively by the conservation low of both
       momentum and energy, because the dedicated detectors at LHC has excellent
       performance on trigger, time resolution, particle identification and so on
       (see \cite{lhc} for details).
 \end{itemize}
 \item From the phenomenological view
 \begin{itemize}
 \item With very high statistics, we can carefully test the various theoretical
       models, precisely determine the CKM elements, and meticulously search for the
       signals of new physics. This requires more accurate theoretical predictions.
       In this paper, we shall study the $B_{c}$ ${\to}$ $B_{q}^{({\ast})}P$, $B_{q}V$
       decays with QCD factorization approach, including the contributions of both
       current-current and penguin operators.
 \item In the rest frame of the $B_{c}$ meson, the velocity ${\beta}_{B_{q}^{({\ast})}}$
       of the $B_{q}^{({\ast})}$ meson is very small due to its large mass, not
       exceeding $0.18$. The ratio of velocity ${\beta}_{P,V}/{\beta}_{B_{q}^{({\ast})}}$
       ${\gtrsim}$ $5.5$, which is very different from that in the two-body $D$ meson
       decays where the ratio of velocities of final states is close to one. This may
       indicate that the final state interferences for $B_{c}$ ${\to}$ $B_{q}^{({\ast})}P$,
       $B_{q}V$ decays might not be so strong as that in $D$ mesons.
       If it holds true, it will benefit us in determining the CKM elements $V_{cs}$ and
       $V_{cd}$, the $B_{c}$ ${\to}$ $B_{q}^{({\ast})}$ transition form factors, etc.
       In this paper, we shall neglect the effects of final state interferences for
       the moment.
 \end{itemize}
 \end{enumerate}

 This paper is organized as follows:
 In Sec. \ref{sec2}, the theoretical framework is discussed. To estimate the amplitude
 of the branching ratios, the master QCD factorization (QCDF) formula are applied to the
 $B_{c}$ ${\to}$ $B_{q}^{({\ast})}P$, $B_{q}V$ decays at the leading approximation.
 Section \ref{sec3} is devoted to the numerical results.
 Finally, we summarize in Sec. \ref{sec4}.

 \section{Theoretical framework}
 \label{sec2}
 \subsection{The effective Hamiltonian}
 \label{sec21}
 Using the operator product expansion and renormalization group (RG) equation,
 the low energy effective Hamiltonian relevant to the $B_{c}$ ${\to}$
 $B_{q}^{({\ast})}P$, $B_{q}V$ decays can be written as
 \begin{equation}
 {\cal H}_{\rm eff} = \frac{G_{F}}{\sqrt{2}} \Big\{ \!\!
  \sum\limits_{q_{1}=d,s\atop q_{2}=d,s} \!\!\! V_{uq_{_{1}}}V_{cq_{_{2}}}^{\ast}
  \Big[ C_{1}({\mu})Q_{1}+ C_{2}({\mu})Q_{2} \Big] + \!\!\!
  \sum\limits_{q=d,s\atop i=3,\ldots,10} \!\!\!\!\! V_{uq}V_{cq}^{\ast}
  C_{i}({\mu})Q_{i} \Big\} + \hbox{H.c.},
 \label{eq:Hamiltonian}
 \end{equation}
 where $V_{uq_{_{\!1}}}V_{cq_{_{2}}}^{\ast}$ is the CKM factor.
 The cases $q$ $=$ $d$ and $q$ $=$ $s$ can be treated separately
 and have the same Wilson coefficients $C_{i}({\mu})$.
  The expressions of the local operators are
 \begin{eqnarray}
 & & Q_{1}=(\bar{u}_{\alpha}q_{\!1\,{\alpha}})_{V-A}(\bar{q}_{2{\beta}}c_{\beta})_{V-A},
      \ \ \ \ \ \ \ \ \ \ \ \ \ \
     Q_{2}=(\bar{u}_{\alpha}q_{1{\beta}})_{V-A}(\bar{q}_{2{\beta}}c_{\alpha})_{V-A}
 \label{eq:operator0102} \\
  & &Q_{3}=({\bar{u}}_{\alpha}c_{\alpha})_{V-A}\sum\limits_{q^{\prime}}
           ({\bar{q}}^{\prime}_{\beta} q^{\prime}_{\beta} )_{V-A},
     \ \ \ \ \ \ \ \ \ \ \ \
     Q_{4}=({\bar{u}}_{\alpha} c_{\beta})_{V-A}\sum\limits_{q^{\prime}}
           ({\bar{q}}^{\prime}_{\beta}q^{\prime}_{\alpha} )_{V-A},
  \label{eq:operator0304} \\
  & &Q_{5}=({\bar{u}}_{\alpha}c_{\alpha})_{V-A}\sum\limits_{q^{\prime}}
           ({\bar{q}}^{\prime}_{\beta} q^{\prime}_{\beta} )_{V+A},
     \ \ \ \ \ \ \ \ \ \ \ \
     Q_{6}=({\bar{u}}_{\alpha} c_{\beta})_{V-A}\sum\limits_{q^{\prime}}
           ({\bar{q}}^{\prime}_{\beta}q^{\prime}_{\alpha} )_{V+A},
  \label{eq:operator0506} \\
  & &Q_{7}=\frac{3}{2}({\bar{u}}_{\alpha}c_{\alpha})_{V-A}
           \sum\limits_{q^{\prime}}e_{q^{\prime}}
           ({\bar{q}}^{\prime}_{\beta} q^{\prime}_{\beta} )_{V+A},
     \ \ \ \ \ \
     Q_{8}=\frac{3}{2}({\bar{u}}_{\alpha} c_{\beta})_{V-A}
           \sum\limits_{q^{\prime}}e_{q^{\prime}}
           ({\bar{q}}^{\prime}_{\beta}q^{\prime}_{\alpha} )_{V+A},
  \label{eq:operator0708} \\
  & &Q_{9}=\frac{3}{2}({\bar{u}}_{\alpha}c_{\alpha})_{V-A}
           \sum\limits_{q^{\prime}}e_{q^{\prime}}
           ({\bar{q}}^{\prime}_{\beta} q^{\prime}_{\beta} )_{V-A},
    \ \ \ \ \ \
    Q_{10}=\frac{3}{2}({\bar{u}}_{\alpha} c_{\beta})_{V-A}
           \sum\limits_{q^{\prime}}e_{q^{\prime}}
           ({\bar{q}}^{\prime}_{\beta}q^{\prime}_{\alpha} )_{V-A},
  \label{eq:operator0910}
 \end{eqnarray}
 where the summation over the repeated color indices (${\alpha}$ and ${\beta}$)
 is understood. The Dirac current
 $(\bar{q}_{1}q_{2})_{V{\pm}A}$ $=$ $\bar{q}_{1}{\gamma}(1{\pm}{\gamma}_{5})q_{2}$.
 $q^{\prime}$ denotes all the active quarks at scale ${\mu}$
 $=$ ${\cal O}(m_{c})$, i.e. $q^{\prime}$ $=$ $u$, $d$, $s$, $c$.
 $e_{q^{\prime}}$ denotes the electric change of the corresponding quark $q^{\prime}$
 in the unit of ${\vert}e{\vert}$, which reflects the electroweak origin of $Q_{7}$,
 ${\cdots}$, $Q_{10}$.
 The current-current operators ($Q_{1}$, $Q_{2}$), QCD penguin operators
 ($Q_{3}$, ${\cdots}$, $Q_{6}$), and electroweak penguin operators ($Q_{7}$,
 ${\cdots}$, $Q_{10}$) form a complete basis set under QCD and QED
 renormalization \cite{rmp68p1125}.

 The effective coupling constants --- Wilson coefficients $C_{i}({\mu})$ ---
 are calculated in perturbative theory at a high scale ${\mu}$ ${\sim}$ $m_{W}$
 and evolved down to a characteristic scale ${\mu}$ ${\sim}$ $m_{c}$ using
 the RG equations.
 The Wilson coefficient functions are given by \cite{rmp68p1125}
 \begin{equation}
 \vec{C}({\mu})=U_{4}({\mu},{\mu}_{b})M({\mu}_{b})U_{5}({\mu}_{b},{\mu}_{W})
 \vec{C}({\mu}_{W})
 \end{equation}
 Here $U_{f}({\mu}_{f},{\mu}_{i})$ is the RG evolution matrix for $f$ active
 flavors, which includes the RG-improved perturbative contribution from the
 initial scale ${\mu}_{i}$ down to the final scale ${\mu}_{f}$. The $M({\mu})$
 is the $10{\times}10$ quark-threshold matching matrix. The corresponding formula
 and expressions can be found in Ref. \cite{rmp68p1125}.
 The Wilson coefficients $C_{i}({\mu})$ have been evaluated to the next-to-leading
 order (NLO). Their numerical values in the naive dimensional regularization (NDR)
 scheme are listed in Table \ref{tab01}.

 \subsection{Hadronic matrix elements within the QCDF framework}
 \label{sec22}
 For the weak decays of hadrons, the short-distance effects are well-known and can
 be calculated in perturbation theory. However, the nonperturbative long-distance
 effects responsible for the hadronization from quarks to hadrons still remain
 obscure in several aspects. But to calculate the exclusive weak decays of the
 $B_{c}$ meson, one needs to evaluate the hadronic matrix elements, i.e., the
 weak current operator sandwiched between the initial state of the $B_{c}$ meson
 and the concerned final states, which is the most difficult theoretical work at
 present. Phenomenologically, these hadronic matrix elements are usually
 parameterized into the product of the decay constants and the transition
 form factors based on the argument of color transparency  and the naive
 factorization scheme (NF) \cite{bsw}.
 A few years ago, Beneke, Buchalla, Neubert, and Sachrajda suggested a QCDF
 formula to compute the hadronic matrix elements in the heavy quark limit,
 combining the hard scattering approach with power counting in $1/m_{Q}$
 \cite{9905312} (here $m_{Q}$ is the mass of heavy quark).
 At leading order in the power series of heavy quark mass expansion, the hadronic
 matrix elements can be factorized into ``non-factorizable'' corrections dominated
 by hard gluon exchange and universal non-perturbative part parameterized by the
 form factors and meson's light cone distribution amplitudes. This promising
 approach has been applied to exclusive two-body nonleptonic $B_{u}$,
 $B_{d}$, $B_{s}$ decays \cite{0108141,prd68p054003,npb657p333}.
 It is found that with appropriate parameters, most of the QCDF's
 predictions are in agreement with the present experimental data.
 In this paper, we would like to apply the QCDF approach to the
 $B_{c}$ ${\to}$ $B^{({\ast})}_{q}P$, $B_{q}V$ decays.

 In the heavy quark limit $m_{c}$ $\gg$ ${\Lambda}_{QCD}$, up to power corrections
 of order of the ${\Lambda}_{QCD}/m_{c}$, using the master QCDF formula,
 the hadronic matrix elements for the $B_{c}$ ${\to}$ $B^{({\ast})}_{q}M$ decays
 ($M$ $=$ $P$ or $V$) can be written as \cite{9905312}
 \begin{equation}
 {\langle}B^{({\ast})}_{q}M{\vert}O_{i}{\vert}B_{c}{\rangle}\ =\
  F^{B_{c}{\to}B^{({\ast})}_{q}} {\int}dz \ H(z) {\Phi}_{M}(z)
 \label{eq:qcdf}
 \end{equation}
 where $F^{B_{c}{\to}B^{({\ast})}_{q}}$ is the transition form factor and
 ${\Phi}_{M}(z)$ is the distribution amplitudes for the meson of $M$, which
 are assumed to be nonperturbative and dominated by the soft contributions.
 The hard-scattering kernels $H(z)$ can be calculated in the perturbative
 theory. For details about the QCDF formula Eq.(\ref{eq:qcdf}), please refer
 to Ref.\cite{9905312}.

 To estimate the branching ratios approximately and to have a sense of the
 order of amplitudes, we shall adopt a rough approximation, i.e. at the
 leading order of ${\alpha}_{s}$. Within this approximation, the hard-scattering
 kernel functions become very simple, $H(z)$ $=$ $1$. That is to say the
 long-distance interactions between $M$ and $B_{c}$-$B^{({\ast})}_{q}$ system
 could be neglected. So the integral of ${\Phi}_{M}(z)$ reduces to the
 normalization condition for the distribution amplitudes.
 Furthermore, according to the arguments of QCDF \cite{9905312}, the hard
 interactions with the spectator are power suppressed in the heavy quark
 limit. 
 Therefore it is not surprisingly to reproduce the result of NF.

 In our paper, the annihilation amplitudes are neglected due to some reasons.
 (1) According to the power counting arguments of QCDF \cite{9905312},
 compared with the leading order contributions to the hard scattering
 kernel, the contributions from annihilation topologies are power suppressed.
 (2) The annihilation amplitudes are suppressed by the CKM elements.
 For the decay modes concerned, the CKM factors in the non-annihilation
 amplitudes are
 $V_{ud}V_{cs}^{\ast}$ ${\sim}$ $1$,
 $V_{us}V_{cs}^{\ast}$ ${\sim}$ ${\lambda}$,
 $V_{ud}V_{cd}^{\ast}$ ${\sim}$ ${\lambda}$ and
 $V_{us}V_{cd}^{\ast}$ ${\sim}$ ${\lambda}^{2}$,
 while the annihilation amplitudes are proportional to the CKM factors of
 $V_{cb}V_{ub}^{\ast}$ ${\sim}$ ${\lambda}^{5}$.

 The explicit expressions of decay amplitudes for $B_{c}$ ${\to}$ $B^{({\ast})}_{q}P$,
 $B_{q}V$ decays are collected in appendix \ref{appendix}. In our paper, we define
 \begin{eqnarray}
 a_{i}&{\equiv}&C_{i}+\frac{1}{N_{c}}C_{i+1}
 \ \ \ \ \ \ \ (i={\rm odd})
 \label{eq:ai01} \\
 a_{i}&{\equiv}&C_{i}+\frac{1}{N_{c}}C_{i-1}
 \ \ \ \ \ \ \ (i={\rm even})
 \label{eq:ai02}
 \end{eqnarray}
 where $i$ runs from $1$ to $10$, $C_{i}$ are the Wilson
 coefficients. $N_{c}$ $=$ $3$ is the color number.

 \section{Numerical results and discussions}
 Within the QCDF approach, the decay amplitudes depend on many input parameters
 including the CKM matrix elements, decay constants, form factors, etc.
 These parameters are discussed and specified below.

 \label{sec3}
 \subsection{The CKM matrix elements}
 \label{sec31}
 We will use the Wolfenstein parameterization. Phenomenologically, it is a popular
 approximation of the CKM matrix in which each elements is expanded as a power
 series in the small parameter ${\lambda}$.
 Up to ${\cal O}({\lambda}^{6})$, the CKM elements can be written as \cite{rmp68p1125}
 \begin{eqnarray}
 V_{ud}&=&1-\frac{1}{2}{\lambda}^{2}-\frac{1}{8}{\lambda}^{4}
        +{\cal O}({\lambda}^{6}) \label{eq:vud} \\
 V_{us}&=&{\lambda}+{\cal O}({\lambda}^{6}) \label{eq:vus} \\
 V_{ub}&=&A{\lambda}^{3}({\rho}-i{\eta})    \label{eq:vub} \\
 V_{cd}&=&-{\lambda}+A^{2}{\lambda}^{5}\Big[\frac{1}{2}-({\rho}+i{\eta})\Big]
          +{\cal O}({\lambda}^{6}) \label{eq:vcd} \\
 V_{cs}&=&1-\frac{1}{2}{\lambda}^{2}-\frac{1}{8}{\lambda}^{4}
           -\frac{1}{2}A^{2}{\lambda}^{4}+{\cal O}({\lambda}^{6}) \label{eq:vcs} \\
 V_{cb}&=&A{\lambda}^{2}+{\cal O}({\lambda}^{6}) \label{eq:vcb} \\
 V_{td}&=&A{\lambda}^{3}\Big[1-({\rho}+i{\eta})\Big(1-\frac{1}{2}{\lambda}^{2}\Big)\Big]
          +{\cal O}({\lambda}^{6}) \label{eq:vtd} \\
 V_{ts}&=&-A{\lambda}^{2}+A{\lambda}^{4}\Big[\frac{1}{2}-({\rho}+i{\eta})\Big]
          +{\cal O}({\lambda}^{6}) \label{eq:vts} \\
 V_{tb}&=&1-\frac{1}{2}A^{2}{\lambda}^{4}+{\cal O}({\lambda}^{6}) \label{eq:vtb}
 \end{eqnarray}
 The global fit for the four independent Wolfenstein parameters gives \cite{pdg2006}
 \begin{equation}
 A=0.818^{+0.007}_{-0.017},\ \ \ \
 {\lambda}=0.2272{\pm}0.0010,\ \ \ \
 \bar{\rho}=0.221^{+0.064}_{-0.028},\ \ \ \
 \bar{\eta}=0.340^{+0.017}_{-0.045}
 \label{eq:ckm02}
 \end{equation}
 where the relationship between (${\rho}$, ${\eta}$) and
 ($\bar{\rho}$, $\bar{\eta}$) is \cite{pdg2006}
 \begin{equation}
 {\rho}+i{\eta}=\frac{\sqrt{1-A^{2}{\lambda}^{4}}(\bar{\rho}+i\bar{\eta})}
     {\sqrt{1-{\lambda}^{4}}[1-A^{2}{\lambda}^{4}(\bar{\rho}+i\bar{\eta})]}
 \label{eq:ckm03}
 \end{equation}
 If not stated otherwise, we shall use their central values for illustration.

 \subsection{Decay constants and form factors}
 \label{sec32}
 In principle, information about the decay constants and transition form factors
 of mesons can be obtained from experiments and/or theoretical estimations. Now
 we specify these parameters. The decay constants $f_{P}$ and $f_{V}$ corresponding
 to the pseudoscalar and vector mesons respectively, are defined by
 \begin{equation}
  {\langle}P(q){\vert}({\bar{q}}_{1}q_{2})_{V-A}{\vert}0{\rangle}
   = - i f_{P} q^{\mu}, \ \ \ \ \ \ \
  {\langle}V(q,{\epsilon}){\vert}({\bar{q}}_{1}q_{2})_{V-A}{\vert}0{\rangle}
   = f_{V} m_{V} {\epsilon}^{{\ast}{\mu}},
   \label{eq:meson01}
 \end{equation}
 where ${\epsilon}^{{\ast}}$ is the polarization vector of the vector meson $V$.
 In this paper, we assume ideal mixing between ${\omega}$ and ${\phi}$ mesons,
 i.e. ${\omega}$ = $(u\bar{u}+d\bar{d})/{\sqrt{2}}$ and ${\phi}$ = $s\bar{s}$.
 In fact, the $B_{c}$ ${\to}$ $B_{u}^{({\ast})}{\phi}$ decays are not
 possible, because the $B_{c}$ meson lies below the threshold of the
 $B_{u}^{({\ast})}{\phi}$ system.
 As to the ${\eta}$ and ${\eta}^{\prime}$ mesons, we take the convention in
 Ref. \cite{prd58p114006}, adopting the Feldmann-Kroll-Stech mixing scheme.
 Neglecting the possible compositions of both ${\eta}_{c}$ $=$ $c\bar{c}$
 and glueball $gg$, the ${\eta}$ and ${\eta}^{\prime}$ are expressed as linear
 combinations of orthogonal states ${\eta}_{q}$ and ${\eta}_{s}$ with the
 flavor structure $q\bar{q}$ $=$ $(u\bar{u}+d\bar{d})/{\sqrt{2}}$
 and $s\bar{s}$, respectively, i.e.
 \begin{equation}
 \left( \begin{array}{c} {\eta} \\ {{\eta}^{\prime}} \end{array} \right) =
 \left( \begin{array}{lr} {\cos}{\phi} & -{\sin}{\phi} \\
       {\sin}{\phi} & {\cos}{\phi} \end{array} \right)
 \left( \begin{array}{c} {\eta}_{q} \\  {\eta}_{s} \end{array} \right)
 \label{eq:meson02}
 \end{equation}
 where ${\phi}$ $=$ $(39.3{\pm}1.0)^{\circ}$ \cite{prd58p114006} is the
 ${\eta}$-${\eta}^{\prime}$ mixing angle. So the decay constants
 related to the ${\eta}$ and ${\eta}^{\prime}$ mesons can be defined by
 \begin{equation}
 \left(\begin{array}{cc} f^{q}_{\eta} & f^{s}_{\eta} \\
     f^{q}_{{\eta}^{\prime}} & f^{s}_{{\eta}^{\prime}} \end{array} \right)
  =  \left(\begin{array}{cc} {\cos}{\phi} & -{\sin}{\phi} \\
     {\sin}{\phi} & {\cos}{\phi} \end{array} \right)
     \left(\begin{array}{cc} f_{q} & 0 \\ 0 & f_{s} \end{array} \right)
 \label{eq:meson03}
 \end{equation}
 \begin{equation}
 {\langle}0{\vert}\bar{q}{\gamma}_{\mu}{\gamma}_{5}q
           {\vert}{\eta}^{(\prime)}(p){\rangle}
          =if^{q}_{{\eta}^{(\prime)}}p_{\mu},
  \ \ \ \ \ \ \
 {\langle}0{\vert}\bar{s}{\gamma}_{\mu}{\gamma}_{5}s
           {\vert}{\eta}^{(\prime)}(p){\rangle}
          =if^{s}_{{\eta}^{(\prime)}}p_{\mu}.
 \label{eq:meson04}
 \end{equation}
 The matrix elements of the pseudoscalar densities are defined by \cite{prd58p094009}
 \begin{equation}
 \frac{{\langle}0{\vert}\bar{u}{\gamma}_{5}u
                 {\vert}{\eta}^{(\prime)}{\rangle}}
      {{\langle}0{\vert}\bar{s}{\gamma}_{5}s
                 {\vert}{\eta}^{(\prime)}{\rangle}}
  = \frac{f^{u}_{{\eta}^{(\prime)}}}{f^{s}_{{\eta}^{(\prime)}}},
    \ \ \ \ \ \ \ \ \ \
       {\langle}0{\vert}\bar{s}{\gamma}_{5}s
                 {\vert}{\eta}^{(\prime)}{\rangle}
  = -i\frac{m_{{\eta}^{(\prime)}}^{2}}{2m_{s}}
   (f^{s}_{{\eta}^{(\prime)}}-f^{u}_{{\eta}^{(\prime)}}),
 \label{eq:meson05}
 \end{equation}
 The numerical values of the decay constants are collected in Table \ref{tab02}.
 If not stated otherwise, we shall take their central values for illustration.

 The transition form factors are defined as \cite{bsw}
 \begin{equation}
  {\langle}P(k){\vert}({\bar{q}}_{3}q_{4})_{V-A}{\vert}B(p){\rangle}=
   (p+k)^{\mu}F_{1}^{B{\to}P}(q^{2})+
   \frac{m_{B}^{2}-m_{P}^{2}}{q^{2}}q^{\mu}
   \Big[F_{0}^{B{\to}P}(q^{2})-F_{1}^{B{\to}P}(q^{2})\Big]
   \label{eq:meson06}
 \end{equation}
 \begin{eqnarray}
 {\langle}V(k,{\epsilon}){\vert}({\bar{q}}_{3}q_{4})_{V-A}{\vert}B(p){\rangle}&=&
 i \frac{{\epsilon}^{\ast}{\cdot}p}{q^{2}}q_{\mu}2m_{V}A_{0}^{B{\to}V}(q^{2}) +
 i{\epsilon}^{\ast}_{\mu}(m_{B}+m_{V}) A_{1}^{B{\to}V}(q^{2}) \nonumber \\ &-&
 i \frac{{\epsilon}^{\ast}{\cdot}p}{m_{B}+m_{V}}(p+k)_{\mu}A_{2}^{B{\to}V}(q^{2}) -
 i \frac{{\epsilon}^{\ast}{\cdot}p}{q^{2}}q_{\mu}2m_{V}A_{3}^{B{\to}V}(q^{2}) \nonumber \\ &+&
  {\epsilon}_{{\mu}{\nu}{\alpha}{\beta}}{\epsilon}^{{\ast}{\nu}}p^{\alpha}k^{\beta}
   \frac{2V^{B{\to}V}(q^{2})}{m_{B}+m_{V}}
 \label{eq:meson07}
 \end{eqnarray}
 where $F_{0,1}$, $V$ and $A_{0,1,2,3}$ are the transition form factors,
 $q$ = $p$ $-$ $k$.
 In order to cancel the poles at $q^{2}$ $=$ $0$, we must impose the condition
 \begin{eqnarray} & &
  F_{0}^{B{\to}P}(0)=F_{1}^{B{\to}P}(0),
  \ \ \ \ \ \ \ \ \ \ \ \ \ \ \ \ \ \
  A_{0}^{B{\to}V}(0)=A_{3}^{B{\to}V}(0),
 \label{eq:meson08} \\ & &
   2 m_{V} A_{3}^{B{\to}V}(0)
 = (m_{B} + m_{V}) A_{1}^{B{\to}V}(0)
 - (m_{B} - m_{V}) A_{2}^{B{\to}V}(0).
 \label{eq:meson09}
 \end{eqnarray}
 In our paper, only the $B_{c}$ ${\to}$ $B^{({\ast})}_{q}$ transition form
 factors appear in the amplitudes within the ``spectator'' model where the
 spectator is the $b$-quark for the concerned processes.
 Their numerical values are collected in Table \ref{tab03}.
 From the numbers in Table \ref{tab03}, we can see clearly that, due to the
 properties of nonperturbative QCD, there are large uncertainties about the
 form factors with different theoretical treatments. Here, we notice the fact
 that the velocity of the final state $B_{q}^{({\ast})}$ meson is very small
 in the rest frame of the initial $B_{c}$ meson, as that mentioned in Sec.
 \ref{sec1}. It is commonly assumed that the velocities of the $b$-quark in
 the rest frame of the $b$-flavored mesons should be close to zero.
 The $B_{q}^{({\ast})}$ meson is neither fast nor small. By intuition, the
 overlap between the initial and final states should be huge, close to unity,
 as that argued in \cite{prd39p1342}. So for illustration and simplification,
 we will take the same value for the transition form factors, i.e.
 $F_{1,0}(0)$ $=$ $A_{0}(0)$ $=$ $0.8$.

 \subsection{Quark masses}
 \label{sec33}
 In the decay amplitudes, there exist the ``chirally enhanced'' factors which
 are associated with the hadronic matrix elements of the scalar and pseudoscalar
 densities, for example, $R_{c1}$ in Eq.(\ref{eq:Bd0pip}). These factors
 are formally of order the ${\Lambda}_{\rm QCD}/m_{c}$, power suppressed in the
 heavy quark limit, but numerically close to unity because the mass of the
 $c$ quark is not infinity in practice.
 The current quark masses in the denominator appear through the equations of
 motions and are renormalization scale dependent. Their values are
 \cite{pdg2006}
 \begin{equation}
 \begin{array}{lll}
    m_{u}(2\,{\rm GeV})=3{\pm}1~{\rm MeV}, &~~~~~
  & m_{d}(2\,{\rm GeV})=6.0{\pm}1.5~{\rm MeV}, \\
    m_{s}(2\,{\rm GeV})=103{\pm}20~{\rm MeV}, &
  & m_{c}(m_{c})=1.24{\pm}0.09~{\rm GeV}.
 \end{array}
 \label{eq:quark01}
 \end{equation}
 Using the renormalization group equation of the running quark mass
 \cite{rmp68p1125},
 \begin{equation}
 m({\mu})=m({\mu}_{0})\Big[\frac{{\alpha}_{s}({\mu})}{{\alpha}_{s}({\mu}_{0})}
    \Big]^{\frac{{\gamma}_{m}^{(0)}}{2{\beta}_{0}}}
    \Big\{1+ \Big(\frac{{\gamma}_{m}^{(1)}}{2{\beta}_{0}}
   -\frac{{\gamma}_{m}^{(0)}{\beta}_{1}}{2{\beta}_{0}^{2}}\Big)
    \frac{{\alpha}_{s}({\mu})-{\alpha}_{s}({\mu}_{0})}{4{\pi}}
    \Big\}
 \label{eq:quark02}
 \end{equation}
 their corresponding values at a characteristic scale ${\mu}$ ${\sim}$ $m_{c}$
 can be obtained.

 \subsection{Numerical results and discussions}
 \label{sec34}
 The numerical results are listed in Table \ref{tab04}, where ${\cal B}r^{T}$
 corresponds to the contributions of the current-current operators only,
 ${\cal B}r^{T+P_{s}}$ corresponds to the contributions of both current-current
 and QCD penguin operators, ${\cal B}r^{T+P_{s}+P_{\rm e}}$ corresponds to the
 contributions of both current-current and penguin operators, i.e. $Q_{1}$,
 ${\cdots}$, $Q_{10}$.

 Here, we would like to point out that these numbers are just the qualitative
 estimations on the order of amplitudes, because many of the subtleties and
 details, such as final state interactions, the renormalization scale dependence,
 the transition form factors, the strong phases, and so on, all deserve the
 dedicated researches but are not considered here.

 From the numbers in Table \ref{tab04}, we can see
 \begin{itemize}
 \item The contributions of both QCD and electroweak penguin operators are
       very small for $B_{c}$ ${\to}$ $B^{({\ast})}P$, $BV$ decays, compared
       with those of the current-current operators. This is very different
       from that of the $B_{u,d,s}$ meson decays. The reason is that the
       contributions of penguin operators are seriously suppressed by the CKM
       elements. The CKM elements corresponding to different topologies for
       $c$-quark decay in the $B_{c}$ meson are listed below.
       \begin{center}
       \begin{tabular}{l|c|c} \hline
       \multicolumn{1}{c|}{tree topologies} &
       penguin topologies & annihilation topologies \\ \hline
       $V_{ud}V_{cs}^{\ast}$ ${\sim}$ $1$,~~~~
       $V_{us}V_{cs}^{\ast}$ ${\sim}$ $+{\lambda}$
     & $V_{ud}V_{cd}^{\ast}$ $+$ $V_{us}V_{cs}^{\ast}$ ${\sim}$ ${\lambda}^{5}$
     & $V_{cb}V_{ub}^{\ast}$ ${\sim}$ ${\lambda}^{5}$  \\
       $V_{us}V_{cd}^{\ast}$ ${\sim}$ ${\lambda}^{2}$,~~~
       $V_{ud}V_{cd}^{\ast}$ ${\sim}$ $-{\lambda}$ & &\\ \hline
       \end{tabular}
       \end{center}
       So, for the $B_{c}$ ${\to}$ $B^{({\ast})}P$, $BV$ decays, the effects of
       new physics contributed via the penguin topologies might be tiny and not
       detectable even with large statistics, due to the serious suppression by
       the CKM elements.
 \item There are clear hierarchy of amplitudes of the branching ratios.
       According the CKM elements and the coefficients of $a_{1,2}$,
       these decay modes are divided into different cases listed below.
       \begin{center}
       \begin{tabular}{l|c|c|c|c} \hline
        cases   & processes & coefficients & the CKM elements & order of branching ratios \\ \hline
        case 1a & $c$ ${\to}$ $s$ & $a_{1}$ & $V_{ud}V_{cs}^{\ast}$ ${\sim}$ $1$ & ${\sim}$ $10^{-2}$ \\
        case 1b & $c$ ${\to}$ $s$ & $a_{1}$ & $V_{us}V_{cs}^{\ast}$ ${\sim}$ ${\lambda}$ & ${\sim}$ $10^{-3}$ \\
                & $c$ ${\to}$ $d$ & $a_{1}$ & $V_{ud}V_{cd}^{\ast}$ ${\sim}$ ${\lambda}$ & ${\sim}$ $10^{-3}$ \\
        case 1c & $c$ ${\to}$ $d$ & $a_{1}$ & $V_{us}V_{cd}^{\ast}$ ${\sim}$ ${\lambda}^{2}$ & ${\sim}$ $10^{-4}$ \\ \hline
        case 2a & $c$ ${\to}$ $u$ & $a_{2}$ & $V_{ud}V_{cs}^{\ast}$ ${\sim}$ $1$ & ${\sim}$ $10^{-5}$ \\
        case 2b & $c$ ${\to}$ $u$ & $a_{2}$ & $V_{us}V_{cs}^{\ast}$,
                            $V_{ud}V_{cd}^{\ast}$ ${\sim}$ ${\lambda}$ & ${\sim}$ $10^{-6}$ --- $10^{-7}$ \\
        case 2c & $c$ ${\to}$ $u$ & $a_{2}$ & $V_{us}V_{cd}^{\ast}$ ${\sim}$ ${\lambda}^{2}$ & ${\sim}$ $10^{-8}$ \\ \hline
       \end{tabular}
       \end{center}
       The decay modes determined by $a_{1}$ have comparatively large branching ratios,
       which should be detectable experimentally, especially the CKM favored decay
       modes, such as $B_{c}$ ${\to}$ $B_{s}^{({\ast})}{\pi}$, $B_{s}{\rho}$, might be
       the promising decay modes to be measured in hadron
       colliders. Due to the great branching ratios of the decay modes determined
       by $a_{1}$, the $B_{c}$ mesons can be used as a source of the $B_{s}$ mesons
       if the $B_{c}$ is produced copiously, as that stated in Ref.\cite{ijmpa21p777}.
       The decay modes determined by $a_{2}$ have comparatively small branching ratios,
       which are hard to detect experimentally, especially the CKM suppressed decay
       modes, such as $B_{c}$ ${\to}$ $B_{u}^{({\ast})}K^{0}$, $B_{u}K^{{\ast}0}$,
       their branching ratios are too tiny to be measured.
 \item Although the $B_{c}$ ${\to}$ $B_{u}^{({\ast})}{\eta}^{\prime}$ decays belong
       to the case 2b modes, their branching ratios are abnormally small, order of
       $10^{-8}$. This can be explained by the fact that on one hand the physical space phase
       available is too small, on the other hand there are large destructive interactions
       between $f^{u}_{{\eta}^{\prime}}a_{2}$ and $f^{s}_{{\eta}^{\prime}}a_{2}$
       due to the serious cancellation between the CKM elements $V_{ud}V_{cd}^{\ast}$
       and $V_{us}V_{cs}^{\ast}$.
 \item The relations among the $B_{c}$ ${\to}$ $B_{q}^{({\ast})}P$, $B_{q}V$ decay mode
       become very simple since the effects of penguin topologies is too tiny to be considered.
       We can use these relations to determine and overconstrain some parameters, such as
       the CKM elements, the form factors, etc. In addition, in estimating and measuring
       these parameters, the ratios of the branching ratios can be used to cancel and/or
       reduce largely theoretical uncertainties and experimental errors.
       For example
       \begin{eqnarray} & &
       \frac{{\cal B}r(B^{+}_{c}{\to}B^{0}_{d}{\pi}^{+})}
            {{\cal B}r(B^{+}_{c}{\to}B^{0}_{d}K^{+})}{\approx}
       \frac{V_{ud}V_{cd}^{\ast}}{V_{us}V_{cd}^{\ast}}
       \frac{f_{\pi}}{f_{K}} {\approx}
       \frac{V_{ud}V_{cs}^{\ast}}{V_{us}V_{cs}^{\ast}}
       \frac{f_{\pi}}{f_{K}} {\approx}
       \frac{{\cal B}r(B^{+}_{c}{\to}B^{0}_{s}{\pi}^{+})}
            {{\cal B}r(B^{+}_{c}{\to}B^{0}_{s}K^{+})}
       \label{eq:ratio01} \\ & &
       \frac{{\cal B}r(B^{+}_{c}{\to}B^{0}_{s}{\pi}^{+})}
            {{\cal B}r(B^{+}_{c}{\to}B^{0}_{d}{\pi}^{+})}{\approx}
       \frac{V_{cd}^{\ast}}{V_{cs}^{\ast}}
       \frac{F_{0}^{B_{c}{\to}B_{d}}(m_{B_{c}}^{2}-m_{B_{d}}^{2})}
            {F_{0}^{B_{c}{\to}B_{s}}(m_{B_{c}}^{2}-m_{B_{s}}^{2})}{\approx}
       \frac{{\cal B}r(B^{+}_{c}{\to}B^{0}_{s}K^{+})}
            {{\cal B}r(B^{+}_{c}{\to}B^{0}_{d}K^{+})}
       \label{eq:ratio02} \\ & &
       \frac{{\cal B}r(B^{+}_{c}{\to}B^{0}_{d}{\pi}^{+})}
            {{\cal B}r(B^{+}_{c}{\to}B^{0}_{d}{\rho}^{+})}{\approx}
       \frac{f_{\pi}F_{0}^{B_{c}{\to}B_{d}}}{f_{\rho}F_{1}^{B_{c}{\to}B_{d}}}
       ~~~~~
       \frac{{\cal B}r(B^{+}_{c}{\to}B^{0}_{d}K^{+})}
            {{\cal B}r(B^{+}_{c}{\to}B^{0}_{d}K^{{\ast}+})}{\approx}
       \frac{f_{K}F_{0}^{B_{c}{\to}B_{d}}}{f_{K^{\ast}}F_{1}^{B_{c}{\to}B_{d}}}
       \label{eq:ratio03} \\ & &
       \frac{{\cal B}r(B^{+}_{c}{\to}B^{0}_{d}{\pi}^{+})}
            {{\cal B}r(B^{+}_{c}{\to}B^{{\ast}0}_{d}{\pi}^{+})}{\approx}
       \frac{F_{0}^{B_{c}{\to}B_{d}}}
            {A_{0}^{B_{c}{\to}B^{\ast}_{d}}} {\approx}
       \frac{{\cal B}r(B^{+}_{c}{\to}B^{0}_{d}K^{+})}
            {{\cal B}r(B^{+}_{c}{\to}B^{{\ast}0}_{d}K^{+})}
       \label{eq:ratio04} \\ & &
       \frac{{\cal B}r(B^{+}_{c}{\to}B^{0}_{s}{\pi}^{+})}
            {{\cal B}r(B^{+}_{c}{\to}B^{0}_{s}{\rho}^{+})}{\approx}
       \frac{f_{\pi}F_{0}^{B_{c}{\to}B_{s}}}{f_{\rho}F_{1}^{B_{c}{\to}B_{s}}}
       ~~~~~
       \frac{{\cal B}r(B^{+}_{c}{\to}B^{0}_{s}K^{+})}
            {{\cal B}r(B^{+}_{c}{\to}B^{0}_{s}K^{{\ast}+})}{\approx}
       \frac{f_{K}F_{0}^{B_{c}{\to}B_{s}}}{f_{K^{\ast}}F_{1}^{B_{c}{\to}B_{s}}}
       \label{eq:ratio05} \\ & &
       \frac{{\cal B}r(B^{+}_{c}{\to}B^{0}_{s}{\pi}^{+})}
            {{\cal B}r(B^{+}_{c}{\to}B^{{\ast}0}_{s}{\pi}^{+})}{\approx}
       \frac{F_{0}^{B_{c}{\to}B_{d}}}
            {A_{0}^{B_{c}{\to}B^{\ast}_{d}}} {\approx}
       \frac{{\cal B}r(B^{+}_{c}{\to}B^{0}_{s}K^{+})}
            {{\cal B}r(B^{+}_{c}{\to}B^{{\ast}0}_{s}K^{+})}
       \label{eq:ratio06}
       \end{eqnarray}
 \end{itemize}

 \section{Summary and Conclusion}
 \label{sec4}
 In prospects of the huge statistics of the $B_{c}$ mesons at the hadron colliders,
 accurate and thorough studies of the $B_{c}$ physics will be accessible very soon.
 In this paper, we study the two-body nonleptonic $c$-quark decays in the $B_{c}$
 mesons, i.e. $B_{c}$ ${\to}$ $B_{q}^{({\ast})}P$, $B_{q}V$ decays within the QCDF
 approach for the leading approximation, and estimate their branching ratios.
 We find that the contributions of the
 penguin operators are very small to the decay amplitudes due to the serious
 suppression by the CKM elements. The decay modes determined by $a_{1}$ have
 comparatively large branching ratios. The most promising decay modes are
 $B_{c}$ ${\to}$ $B_{s}^{({\ast})}{\pi}$, $B_{s}{\rho}$, which might be easily
 detected at the hadron colliders.

 \begin{appendix}

 \section{Amplitudes for $B_{c}$ ${\to}$ $B_{q}^{({\ast})}P$, $B_{q}V$ decays}
 \label{appendix}
 \subsection{$c$ ${\to}$ $d$ processes}
 \begin{eqnarray}
 {\cal A}(B_{c}^{+}{\to}B_{d}^{0}{\pi}^{+}) &=&
 -i\frac{G_{F}}{\sqrt{2}} f_{\pi} F_{0}^{B_{c}{\to}B_{d}^{0}}
  \left(m_{B_{c}}^{2} - m_{B_{d}}^{2}\right) \Big\{
  V_{ud}V_{cd}^{\ast} a_{1} \nonumber \\ & & +
  \Big( V_{ud}V_{cd}^{\ast} + V_{us}V_{cs}^{\ast} \Big)
  \Big[ a_{4} - \frac{1}{2} a_{10} +
  R_{c1} \big( a_{6} - \frac{1}{2} a_{8} \big) \Big] \Big\}
 \label{eq:Bd0pip}
 \end{eqnarray}
 where
 \begin{math}
 R_{c1}\ =\ \displaystyle \frac{2m_{{\pi}^{+}}^{2}}{(m_{d}+m_{u})(m_{c}-m_{d})}
 \end{math}.
 \begin{equation}
 {\cal A}(B_{c}^{+}{\to}B_{d}^{0}K^{+}) =
 -i\frac{G_{F}}{\sqrt{2}} f_{K} F_{0}^{B_{c}{\to}B_{d}^{0}}
  \left(m_{B_{c}}^{2}-m_{B_{d}}^{2}\right) V_{us}V_{cd}^{\ast} a_{1}
 \label{eq:Bd0kp}
 \end{equation}
 \begin{eqnarray}
 {\cal A}(B_{c}^{+}{\to}B_{d}^{0}{\rho}^{+}) &=&
  \sqrt{2}G_{F}f_{{\rho}} F_{1}^{B_{c}{\to}B_{d}^{0}}
  m_{{\rho}^{+}} \left({\varepsilon}{\cdot}p_{_{B_{c}}}\right)
  \Big\{  V_{ud}V_{cd}^{\ast} a_{1} \nonumber \\ & &+
  \Big( V_{ud}V_{cd}^{\ast} + V_{us}V_{cs}^{\ast} \Big) \Big[
  a_{4} - \frac{1}{2} a_{10} \Big] \Big\}
  \label{eq:Bd0rhop}
 \end{eqnarray}
  \begin{equation}
 {\cal A}(B_{c}^{+}{\to}B_{d}^{0}K^{{\ast}+}) =
  \sqrt{2}G_{F}f_{K^{\ast}} F_{1}^{B_{c}{\to}B_{d}^{0}} m_{K^{{\ast}+}}
  \left({\varepsilon}{\cdot}p_{_{B_{c}}}\right) V_{us}V_{cd}^{\ast} a_{1}
  \label{eq:Bd0vkp}
 \end{equation}
 \begin{eqnarray}
 {\cal A}(B_{c}^{+}{\to}B_{d}^{{\ast}0}{\pi}^{+}) &=&
  \sqrt{2}G_{F} f_{\pi} A_{0}^{B_{c}{\to}B_{d}^{{\ast}0}}
  m_{B_{d}^{{\ast}}} \left({\varepsilon}{\cdot}p_{_{B_{c}}} \right)
  \Big\{ V_{ud}V_{cd}^{\ast} a_{1} \nonumber \\ & & +
  \Big( V_{ud}V_{cd}^{\ast} + V_{us}V_{cs}^{\ast} \Big) \Big[
  a_{4} - \frac{1}{2} a_{10} + Q_{c1} \big( a_{6} -
  \frac{1}{2} a_{8} \big) \Big] \Big\}
  \label{eq:vBd0pip}
 \end{eqnarray}
 where
 \begin{math}
 Q_{c1}\ =\ \displaystyle \frac{-2m_{{\pi}^{+}}^{2}}{(m_{d}+m_{u})(m_{c}+m_{d})}
 \end{math}.
 \begin{equation}
 {\cal A}(B_{c}^{+}{\to}B_{d}^{{\ast}0}K^{+}) =
  \sqrt{2}G_{F}f_{K} A_{0}^{B_{c}{\to}B_{d}^{{\ast}0}}
  m_{B_{d}^{{\ast}}} \left({\varepsilon}{\cdot}p_{_{B_{c}}} \right)
  V_{us}V_{cd}^{\ast} a_{1}
 \label{eq:vBd0kp}
 \end{equation}

 \subsection{$c$ ${\to}$ $s$ processes}
 \begin{equation}
 {\cal A}(B_{c}^{+}{\to}B_{s}^{0}{\pi}^{+})\ =\
 -i\frac{G_{F}}{\sqrt{2}} f_{\pi} F_{0}^{B_{c}{\to}B_{s}^{0}}
  \left(m_{B_{c}}^{2}-m_{B_{s}}^{2}\right) V_{ud}V_{cs}^{\ast} a_{1}
 \label{eq:Bs0pip}
 \end{equation}
 \begin{eqnarray}
 {\cal A}(B_{c}^{+}{\to}B_{s}^{0}K^{+})  &=&
 -i\frac{G_{F}}{\sqrt{2}} f_{K} F_{0}^{B_{c}{\to}B_{s}^{0}}
  \left(m_{B_{c}}^{2} - m_{B_{s}}^{2} \right)
  \Big\{ V_{us}V_{cs}^{\ast} a_{1} \nonumber \\ & & +
  \Big( V_{ud}V_{cd}^{\ast} + V_{us}V_{cs}^{\ast} \Big) \Big[
  a_{4} - \frac{1}{2} a_{10} + R_{c2} \big( a_{6} -
  \frac{1}{2} a_{8} \big) \Big] \Big\}
 \label{eq:Bs0kp}
 \end{eqnarray}
 where
 \begin{math}
 R_{c2}\ =\ \displaystyle\frac{2m_{K^{+}}^{2}}{(m_{s}+m_{u})(m_{c}-m_{s})}
 \end{math}
 \begin{equation}
 {\cal A}(B_{c}^{+}{\to}B_{s}^{0}{\rho}^{+})\ =\
  \sqrt{2}G_{F}f_{{\rho}} F_{1}^{B_{c}{\to}B_{s}^{0}} m_{{\rho}^{+}}
  \left({\varepsilon}{\cdot}p_{_{B_{c}}} \right) V_{ud}V_{cs}^{\ast} a_{1}
 \label{eq:Bs0rhop}
 \end{equation}
 \begin{eqnarray}
 {\cal A}(B_{c}^{+}{\to}B_{s}^{0}K^{{\ast}+}) &=&
  \sqrt{2}G_{F}f_{K^{{\ast}}} F_{1}^{B_{c}{\to}B_{s}^{0}} m_{K^{{\ast}+}}
  \left({\varepsilon}{\cdot}p_{_{B_{c}}}\right)
  \Big\{  V_{us}V_{cs}^{\ast} a_{1} \nonumber \\ & & +
  \Big(  V_{ud}V_{cd}^{\ast} + V_{us}V_{cs}^{\ast} \Big) \Big[
  a_{4} - \frac{1}{2} a_{10} \Big] \Big\}
 \label{eq:Bs0vkp}
 \end{eqnarray}
 \begin{equation}
 {\cal A}(B_{c}^{+}{\to}B_{s}^{{\ast}0}{\pi}^{+}) =
  \sqrt{2}G_{F}f_{\pi} A_{0}^{B_{c}{\to}B_{s}^{{\ast}0}} m_{B_{s}^{{\ast}}}
 \left({\varepsilon}{\cdot}p_{_{B_{c}}}\right) V_{ud}V_{cs}^{\ast} a_{1}
 \label{eq:vBs0pip}
 \end{equation}
 \begin{eqnarray}
 {\cal A}(B_{c}^{+}{\to}B_{s}^{{\ast}0}K^{+}) &=&
  \sqrt{2}G_{F} f_{K} A_{0}^{B_{c}{\to}B_{s}^{{\ast}0}} m_{B_{s}^{{\ast}}}
  \left({\varepsilon}{\cdot}p_{_{B_{c}}} \right)
  \Big\{ V_{us}V_{cs}^{\ast} a_{1} \nonumber \\ & & +
  \Big(  V_{ud}V_{cd}^{\ast} + V_{us}V_{cs}^{\ast} \Big) \Big[
  a_{4} - \frac{1}{2} a_{10} + Q_{c2} \big( a_{6} -
  \frac{1}{2} a_{8} \big) \Big] \Big\}
 \label{eq:vBs0kp}
 \end{eqnarray}
 where
 \begin{math}
 Q_{c2}\ =\ \displaystyle \frac{-2m_{K^{+}}^{2}}{(m_{s}+m_{u})(m_{c}+m_{s})}
 \end{math}

 \subsection{$c$ ${\to}$ $u$ processes}
 \begin{eqnarray}
 \lefteqn{ {\cal A}(B_{c}^{+}{\to}B_{u}^{+}{\pi}^{0}) =
 -i\frac{G_{F}}{2}f_{\pi} F_{0}^{B_{c}{\to}B_{u}^{+}}
  \left(m_{B_{c}}^{2}-m_{B_{u}}^{2}\right)
  \Big\{  -V_{ud}V_{cd}^{\ast} a_{2} } \nonumber \\ & & +
  \Big( V_{ud}V_{cd}^{\ast} + V_{us}V_{cs}^{\ast} \Big) \Big[
  a_{4} + a_{10} - \frac{3}{2} \big( a_{7} - a_{9} \big)
  + R_{c3} \big(  a_{6} + a_{8} \big) \Big] \Big\}
 \label{eq:Buppi0}
 \end{eqnarray}
 where
 \begin{math}
 R_{c3}\ =\ \displaystyle\frac{2m_{{\pi}^{0}}^{2}}{(m_{d}+m_{u})(m_{c}-m_{u})}
 \end{math}
 \begin{equation}
 {\cal A}(B_{c}^{+}{\to}B_{u}^{+}{\overline{K}}^{0}) =
 -i\frac{G_{F}}{\sqrt{2}}f_{K} F_{0}^{B_{c}{\to}B_{u}^{+}}
  \left(m_{B_{c}}^{2}-m_{B_{u}}^{2}\right) V_{ud}V_{cs}^{\ast} a_{2}
 \label{eq:Bupkbar0}
 \end{equation}
 \begin{equation}
 {\cal A}(B_{c}^{+}{\to}B_{u}^{+}K^{0}) =
 -i\frac{G_{F}}{\sqrt{2}}f_{K} F_{0}^{B_{c}{\to}B_{u}^{+}}
  \left(m_{B_{c}}^{2}-m_{B_{u}}^{2}\right) V_{us}V_{cd}^{\ast} a_{2}
  \label{eq:Bupk0}
 \end{equation}
 \begin{eqnarray}
 \lefteqn{ {\cal A}(B_{c}^{+}{\to}B_{u}^{+}{\eta}^{(\prime)}) =
 -i\frac{G_{F}}{\sqrt{2}} f_{{\eta}^{(\prime)}}^{u} F_{0}^{B_{c}{\to}B_{u}^{+}}
  \left(m_{B_{c}}^{2}-m_{B_{u}}^{2}\right)
  \Big\{  V_{ud}V_{cd}^{\ast} a_{2} } \nonumber \\ & & +
  \frac{f^{s}_{{\eta}^{(\prime)}}}{f^{u}_{{\eta}^{(\prime)}}}
  V_{us}V_{cs}^{\ast} a_{2}
  + \Big( V_{ud}V_{cd}^{\ast} + V_{us}V_{cs}^{\ast} \Big)
    \Big[ \frac{f^{s}_{{\eta}^{(\prime)}}}{f^{u}_{{\eta}^{(\prime)}}} \big\{
    a_{3} - a_{5} + \frac{1}{2} \Big( a_{7} - a_{9} \Big) \big\}
  \nonumber \\ & & + 2 \Big( a_{3} - a_{5} \Big)
  + a_{4} + a_{10} - \frac{1}{2} \Big( a_{7} - a_{9} \Big)
  + \left(1 - \frac{f^{u}_{{\eta}^{(\prime)}}}{f^{s}_{{\eta}^{(\prime)}}}\right)
    R_{c4}^{(\prime)} \Big( a_{6} + a_{8} \Big) \Big] \Big\}
 \label{eq:Bupeta}
 \end{eqnarray}
 where
 \begin{math}
 R_{c4}^{(\prime)}\ =\
 \displaystyle\frac{2m_{{{\eta}^{(\prime)}}}^{2}}{(m_{s}+m_{s})(m_{c}-m_{u})}
 \end{math}
  \begin{eqnarray}
 {\cal A}(B_{c}^{+}{\to}B_{u}^{+}{\rho}^{0}) &=&
  G_{F}f_{{\rho}} F_{1}^{B_{c}{\to}B_{u}^{+}} m_{{\rho}^{0}}
  \left({\varepsilon}{\cdot}p_{_{B_{c}}}\right)
  \Big\{ - V_{ud}V_{cd}^{\ast} a_{2} \nonumber \\ & & +
  \Big( V_{ud}V_{cd}^{\ast} + V_{us}V_{cs}^{\ast} \Big) \Big[
   a_{4} + a_{10} + \frac{3}{2} \big( a_{7} + a_{9} \big) \Big] \Big\}
 \label{eq:Buprho0}
 \end{eqnarray}
 \begin{eqnarray}
 \lefteqn{ {\cal A}(B_{c}^{+}{\to}B_{u}^{+}{\omega}) =
  G_{F} f_{\omega} F_{1}^{B_{c}{\to}B_{u}^{+}} m_{\omega}
  \left({\varepsilon}{\cdot}p_{_{B_{c}}}\right)
  \Big\{ V_{ud}V_{cd}^{\ast} a_{2} } \nonumber \\ & & +
  \Big( V_{ud}V_{cd}^{\ast} + V_{us}V_{cs}^{\ast} \Big) \Big[
  2 \big( a_{3} + a_{5} \big) + a_{4} + a_{10} + \frac{1}{2}
  \big( a_{7} + a_{9} \big) \Big] \Big\}
 \label{eq:Bupomega}
 \end{eqnarray}
 \begin{equation}
 {\cal A}(B_{c}^{+}{\to}B_{u}^{+}{\overline{K}}^{{\ast}0}) =
 {\sqrt{2}}G_{F} f_{{\overline{K}}^{{\ast}}} F_{1}^{B_{c}{\to}B_{u}^{+}}
 m_{{\overline{K}}^{{\ast}0}} \left({\varepsilon}{\cdot}p_{_{B_{c}}}\right)
 V_{ud}V_{cs}^{\ast} a_{2}
 \label{eq:Bupvkbar0}
 \end{equation}
 \begin{equation}
 {\cal A}(B_{c}^{+}{\to}B_{u}^{+}K^{{\ast}0}) =
 {\sqrt{2}}G_{F} f_{K^{{\ast}0}} F_{1}^{B_{c}{\to}B_{u}^{+}} m_{K^{{\ast}0}}
  \left({\varepsilon}{\cdot}p_{_{B_{c}}}\right) V_{us}V_{cd}^{\ast} a_{2}
 \label{eq:Bupvk0}
 \end{equation}
 \begin{eqnarray}
 \lefteqn{ {\cal A}(B_{c}^{+}{\to}B_{u}^{{\ast}+}{\pi}^{0}) =
 G_{F}f_{\pi} A_{0}^{B_{c}{\to}B_{u}^{{\ast}+}} m_{B_{u}^{{\ast}}}
  \left({\varepsilon}{\cdot}p_{_{B_{c}}}\right)
  \Big\{  -V_{ud}V_{cd}^{\ast} a_{2} } \nonumber \\ & & +
  \Big( V_{ud}V_{cd}^{\ast} + V_{us}V_{cs}^{\ast} \Big) \Big[
  a_{4} + a_{10} - \frac{3}{2} \big( a_{7} - a_{9} \big)
  + Q_{c3} \big(  a_{6} + a_{8} \big) \Big] \Big\}
 \label{eq:vBuppi0}
 \end{eqnarray}
 where
 \begin{math}
 Q_{c3}\ =\ \displaystyle\frac{-2m_{{\pi}^{0}}^{2}}{(m_{d}+m_{u})(m_{c}+m_{u})}
 \end{math}
 \begin{equation}
 {\cal A}(B_{c}^{+}{\to}B_{u}^{{\ast}+}{\overline{K}}^{0}) =
 {\sqrt{2}}G_{F} f_{K} A_{0}^{B_{c}{\to}B_{u}^{{\ast}+}} m_{B_{u}^{{\ast}+}}
  \left({\varepsilon}{\cdot}p_{_{B_{c}}}\right) V_{ud}V_{cs}^{\ast} a_{2}
  \label{eq:vBupkbar0}
 \end{equation}
 \begin{equation}
 {\cal A}(B_{c}^{+}{\to}B_{u}^{{\ast}+}K^{0}) =
 {\sqrt{2}}G_{F} f_{K} A_{0}^{B_{c}{\to}B_{u}^{{\ast}+}} m_{B_{u}^{{\ast}+}}
  \left({\varepsilon}{\cdot}p_{_{B_{c}}}\right) V_{us}V_{cd}^{\ast} a_{2}
 \label{eq:vBupk0}
 \end{equation}
 \begin{eqnarray}
 \lefteqn{ {\cal A}(B_{c}^{+}{\to}B_{u}^{{\ast}+}{{\eta}^{(\prime)}}) =
 {\sqrt{2}}G_{F}f_{{\eta}^{(\prime)}}^{u} A_{0}^{B_{c}{\to}B_{u}^{{\ast}+}}
  m_{B_{u}^{{\ast}+}} \left({\varepsilon}{\cdot}p_{_{B_{c}}}\right)
  \Big\{  V_{ud}V_{cd}^{\ast} a_{2} } \nonumber \\ & & +
  \frac{f^{s}_{{\eta}^{(\prime)}}}{f^{u}_{{\eta}^{(\prime)}}}
   V_{us}V_{cs}^{\ast} a_{2}
  + \Big( V_{ud}V_{cd}^{\ast} + V_{us}V_{cs}^{\ast} \Big)
    \Big[ \frac{f^{s}_{{\eta}^{(\prime)}}}{f^{u}_{{\eta}^{(\prime)}}} \big\{
    a_{3} - a_{5} + \frac{1}{2} \Big( a_{7} - a_{9} \Big) \big\}
  \nonumber \\ & & + 2 \Big( a_{3} - a_{5} \Big)
  + a_{4} + a_{10} - \frac{1}{2} \Big( a_{7}  - a_{9} \Big)
  + \left(1 - \frac{f^{u}_{{\eta}^{(\prime)}}}{f^{s}_{{\eta}^{(\prime)}}}\right)
  Q_{c4}^{({\prime})} \Big( a_{6} + a_{8} \Big) \Big] \Big\}
 \label{eq:vBupeta}
 \end{eqnarray}
 where
 \begin{math}
 Q_{c4}^{({\prime})}\ =\ \displaystyle
 \frac{-2m_{{{\eta}^{(\prime)}}}^{2}}{(m_{s}+m_{s})(m_{c}+m_{u})}
 \end{math}

 \end{appendix}

 \section*{Acknowledgments}
 This work is supported in part both by National Natural Science Foundation
 of China (under Grant No. 10647119, 10710146) and by Natural Science Foundation
 of Henan Province, China. We would like to thank Prof. Dongsheng Du,
 Dr. Deshan Yang, Prof. Caidian L\"{u} and Prof. Zhizhong Xing for valuable discussions.

 \begin{table}[ht]
 \begin{center}
 \caption{The NLO Wilson coefficients $C_{i}(\mu)$ in the NDR scheme.
  The input parameters are \cite{pdg2006}:
  ${\alpha}_{s}(m_{Z})$ $=$ $0.1176$,
  ${\alpha}_{em}(m_{W})$ $=$ $1/128$,
  $m_{W}$ $=$ $80.403$ GeV,
  ${\Lambda}_{\rm QCD}^{(f=5)}$ $=$ $220.9$ MeV,
  ${\Lambda}_{\rm QCD}^{(f=4)}$ $=$ $317.2$ MeV.}
 \label{tab01}
 \begin{ruledtabular}
 \begin{tabular}{c r rrr}
  & \multicolumn{1}{c}{${\mu}=m_{b}$}
  & \multicolumn{1}{c}{${\mu}=2.0$ GeV}
  & \multicolumn{1}{c}{${\mu}=1.5$ GeV}
  & \multicolumn{1}{c}{${\mu}=m_{c}$} \\  \hline
  $C_{1}$                & $ 1.0849$ & $ 1.1497$ & $ 1.1883$ & $ 1.2215$ \\
  $C_{2}$                & $-0.1902$ & $-0.3077$ & $-0.3717$ & $-0.4241$ \\
  $C_{3}$                & $ 0.0148$ & $ 0.0238$ & $ 0.0296$ & $ 0.0349$ \\
  $C_{4}$                & $-0.0362$ & $-0.0542$ & $-0.0652$ & $-0.0747$ \\
  $C_{5}$                & $ 0.0088$ & $ 0.0105$ & $ 0.0107$ & $ 0.0102$ \\
  $C_{6}$                & $-0.0422$ & $-0.0703$ & $-0.0896$ & $-0.1078$ \\
  $C_{7}/{\alpha}_{em}$  & $-0.0007$ & $-0.0164$ & $-0.0186$ & $-0.0181$ \\
  $C_{8}/{\alpha}_{em}$  & $ 0.0565$ & $ 0.0964$ & $ 0.1235$ & $ 0.1493$ \\
  $C_{9}/{\alpha}_{em}$  & $-1.3039$ & $-1.3966$ & $-1.4473$ & $-1.4901$ \\
  $C_{10}/{\alpha}_{em}$ & $ 0.2700$ & $ 0.4144$ & $ 0.4964$ & $ 0.5656$
  \end{tabular}
  \end{ruledtabular}
  \end{center}
  \end{table}

 \begin{table}[ht]
 \caption{values of the decay constant (in the unit of MeV)}
 \label{tab02}
 \begin{ruledtabular}
 \begin{tabular}{cccc|ccc}
 $f_{\pi}$ & $f_{K}$ & $f_{q}$ & $f_{s}$ &
 $f_{\rho}$ & $f_{\omega}$ & $f_{K^{\ast}}$ \\ \hline
 $131$ \cite{pdg2006} & $160$ \cite{pdg2006} &
 $(1.07{\pm}0.02)f_{\pi}$ \cite{prd58p114006} &
 $(1.34{\pm}0.06)f_{\pi}$ \cite{prd58p114006} &
 $205{\pm}9$ \cite{prd71p014029} &
 $195{\pm}3$ \cite{prd71p014029} &
 $217{\pm}5$ \cite{prd71p014029}
 \end{tabular}
 \end{ruledtabular}
 \end{table}

 \begin{table}[ht]
 \caption{Values of transition form factors}
 \label{tab03}
 \begin{ruledtabular}
 \begin{tabular}{c|c|c|c|c}
 Ref.  & $F^{B_{c}{\to}B_{u,d}}_{0}(0)$
       & $F^{B_{c}{\to}B_{s}}_{0}(0)$
       & $A^{B_{c}{\to}B_{u,d}^{\ast}}_{0}(0)$
       & $A^{B_{c}{\to}B_{s}^{\ast}}_{0}(0)$ \\ \hline
 \cite{prd39p1342} \footnotemark[1]
   & $0.320{\sim}0.910$
   & $0.340{\sim}0.925$
   & $0.349{\sim}0.916$
   & $0.432{\sim}0.931$ \\
 \cite{zpc57p43} \footnotemark[2]
   & $0.3{\pm}0.1$
   & $0.30{\pm}0.05$
   & $0.35{\pm}0.09$
   & $0.39{\pm}0.05$ \\
 \cite{9504319} \footnotemark[3]
   & ---  & $0.61$
   & ---  & $0.79$ \\
 \cite{9810339} \footnotemark[4]
   & ---  & $0.403{\sim}0.617$
   & ---  & $0.433{\sim}0.641$ \\
 \cite{jpg26p1079} \footnotemark[5]
   & $0.4504$
   & $0.5917$
   & $0.2691$
   & $0.4451$ \\
 \cite{prd63p074010} \footnotemark[6]
   & $-0.58$
   & $-0.61$
   & $0.35$
   & $0.39$ \\
 \cite{mpla16p1439} \footnotemark[7]
   & ---   & $0.297$
   & ---   & $0.263$ \\
 \cite{pan64p1860} \footnotemark[8]
   & $1.27$
   & $1.3$
   & $1.29$
   & $0.94$ \\
 \cite{pan64p1860} \footnotemark[9]
   & $1.38$
   & $1.1$
   & $1.26$
   & $1.04$ \\
 \cite{epjc32p29}
   & $0.39$
   & $0.50$
   & $0.20$
   & $0.35$ \\
 \cite{epjc47p413}
   & ---
   & ---
   & $0.23{\pm}0.03$
   & --- \\
 \cite{epjc51p833} \footnotemark[10]
   & $0.90$
   & $1.02$
   & $0.27$
   & $0.36$ \\
 \end{tabular}
 \end{ruledtabular}
 \end{table}
 \setcounter{equation}{38}
 \footnotetext[1]{The form factors increase with the increasing parameter
 ${\omega}$ $=$ $0.4$ ${\sim}$ $1.0$ GeV that determines the average
 transverse quark momentum. The authors of \cite{prd39p1342} prefer
 $F^{B_{c}{\to}B_{u}}_{0}(0)$ $=$ $0.831$,
 $F^{B_{c}{\to}B_{s}}_{0}(0)$ $=$ $0.859$,
 $A^{B_{c}{\to}B_{u}^{\ast}}_{0}(0)$ $=$ $0.869$ and
 $A^{B_{c}{\to}B_{s}^{\ast}}_{0}(0)$ $=$ $0.842$
 with the corresponding parameter ${\omega}$ $=$ $0.8$ GeV.}
 \footnotetext[2]{The definitions of the transition form factors in
 \cite{zpc57p43} are different from ours in Eq.(\ref{eq:meson03})
 and Eq.(\ref{eq:meson04}). The relationship is
 \begin{equation}
 F_{1}^{B_{c}{\to}P}=F_{+}, \ \ \ \ \ \ \ \
 A_{0}^{B_{c}{\to}V}=\frac{F_{0}^{A}}{2m_{V}}+
  \frac{m_{_{B_{c}}}^{2}-m_{V}^{2}}{2m_{V}}F_{+}^{A}
 \label{eq:xxx}.
 \end{equation}
 with the values of $F_{+}$ $=$ $0.3{\pm}0.1$ ($0.30{\pm}0.05$),
 $F_{0}^{A}$ $=$ $4.0{\pm}1.0$ ($4.5{\pm}0.5$)  ${\rm GeV}^{-1}$and
 $F_{+}^{A}$ $=$ $-0.02{\pm}0.01$ ($-0.03{\pm}0.02$) ${\rm GeV}^{-1}$
 for $B_{c}$ ${\to}$ $B_{u,d}^{({\ast})}$ ($B_{s}^{({\ast})}$)
 transition \cite{zpc57p43}.}
 \footnotetext[3]{Using the relationship of Eq.(\ref{eq:meson08})
 and Eq.(\ref{eq:meson09}) with the input $A_{1}$ $=$ $0.52$,
 $A_{2}$ $=$ $-2.79$ \cite{9504319}.}
 \footnotetext[4]{For parameter ${\omega}$ $=$ $0.4$, $0.5$ GeV.}
 \footnotetext[5]{Using the relationship of Eq.(\ref{eq:xxx})
 with the input $F_{+}$ $=$ $0.4504$ ($0.5917$),
 $F_{0}^{A}$ $=$ $3.383$ ($5.506$)  ${\rm GeV}^{-1}$ and
 $F_{+}^{A}$ $=$ $-0.0463$ ($-0.0673$)  ${\rm GeV}^{-1}$ for
 $B_{c}$ ${\to}$ $B_{u,d}^{({\ast})}$ ($B_{s}^{({\ast})}$)
 transition \cite{jpg26p1079}.}
 \footnotetext[6]{Using the relationship of Eq.(\ref{eq:meson08})
 and Eq.(\ref{eq:meson09}) with the input $A_{1}$ $=$ $0.27$ ($0.33$)
 and $A_{2}$ $=$ $-0.60$ ($-0.40$) for
 $B_{c}$ ${\to}$ $B_{u,d}^{({\ast})}$ ($B_{s}^{({\ast})}$)
 transition \cite{prd63p074010}.}
 \footnotetext[7]{Using the relationship of Eq.(\ref{eq:meson08})
 and Eq.(\ref{eq:meson09}) with the input $A_{1}$ $=$ $0.28$
 and $A_{2}$ $=$ $0.49$ \cite{mpla16p1439}.}
 \footnotetext[8]{Using the relationship of Eq.(\ref{eq:xxx})
 with the input $F_{+}$ $=$ $1.27$ ($1.3$),
 $F_{0}^{A}$ $=$ $9.8$ ($8.1$)  ${\rm GeV}^{-1}$ and
 $F_{+}^{A}$ $=$ $0.35$ ($0.2$)  ${\rm GeV}^{-1}$ for
 $B_{c}$ ${\to}$ $B_{u,d}^{({\ast})}$ ($B_{s}^{({\ast})}$)
 transition in the framework of QCD sum rules \cite{pan64p1860}.}
 \footnotetext[9]{Using the relationship of Eq.(\ref{eq:xxx})
 with the input $F_{+}$ $=$ $1.38$ ($1.1$),
 $F_{0}^{A}$ $=$ $9.4$ ($8.2$) ${\rm GeV}^{-1}$ and
 $F_{+}^{A}$ $=$ $0.36$ ($0.3$) ${\rm GeV}^{-1}$ for
 $B_{c}$ ${\to}$ $B_{u,d}^{({\ast})}$ ($B_{s}^{({\ast})}$)
 transition in the framework of potential model \cite{pan64p1860}.}
 \footnotetext[10]{Using the relationship of Eq.(\ref{eq:meson08})
 and Eq.(\ref{eq:meson09}) with the input $A_{1}$ $=$ $0.90$ ($1.01$)
 and $A_{2}$ $=$ $7.9$ ($9.04$) for
 $B_{c}$ ${\to}$ $B_{u,d}^{({\ast})}$ ($B_{s}^{({\ast})}$)
 transition \cite{epjc51p833}.}

 \begin{table}[ht]
 \begin{center}
 \caption{The branching ratios for $B_{c}$ ${\to}$ $B_{q}^{({\ast})}P$,
 $B_{q}V$. ${\cal B}r^{T}$ corresponds to the contributions of the
 operators $Q_{1}$ and $Q_{2}$. ${\cal B}r^{T+P_{s}}$ corresponds to
 the contributions of operators $Q_{1}$ ${\sim}$ $Q_{6}$.
 ${\cal B}r^{T+P_{s}+P_{\rm e}}$ corresponds to the contributions of
 $Q_{1}$ ${\sim}$ $Q_{10}$.}
 \label{tab04}
 \begin{ruledtabular}
 \begin{tabular}{lcrrrrr}
    \multicolumn{1}{c}{modes}
  & \multicolumn{1}{c}{case}
  & \multicolumn{1}{c}{${\cal B}r^{T}$}
  & \multicolumn{1}{c}{${\cal B}r^{T+P_{s}}$}
  & \multicolumn{1}{c}{${\cal B}r^{T+P_{s}+P_{\rm e}}$}
  & \multicolumn{1}{c}{$\frac{{\cal B}r^{T+P_{\rm s}}-{\cal B}r^{T}}{{\cal B}r^{T}}$}
  & \multicolumn{1}{c}{$\frac{{\cal B}r^{T+P_{\rm s}+P_{\rm e}}-{\cal B}r^{T}}{{\cal B}r^{T}}$} \\  \hline
 $B_{c}^{+}$ ${\to}$ $B_{s}^{0}{\pi}^{+}$ & case 1a &
 $  5.3089{\times}10^{-2}$ &
  --- & --- & --- & --- \\
  $B_{c}^{+}$ ${\to}$ $B_{s}^{0}{\rho}^{+}$ & case 1a &
 $  6.2652{\times}10^{-2}$ &
  --- & --- & --- & --- \\
  $B_{c}^{+}$ ${\to}$ $B_{s}^{{\ast}0}{\pi}^{+}$ & case 1a &
 $  4.5916{\times}10^{-2}$ &
  --- & --- & --- & --- \\
 $B_{c}^{+}$ ${\to}$ $B_{s}^{0}{K}^{+}$ & case 1b &
 $  3.6746{\times}10^{-3}$ &
 $  3.6759{\times}10^{-3}$ &
 $  3.6759{\times}10^{-3}$ &
 $  3.4{\times}10^{-4}$ &
 $  3.4{\times}10^{-4}$ \\
  $B_{c}^{+}$ ${\to}$ $B_{s}^{0}{K}^{{\ast}+}$ & case 1b &
 $  1.6450{\times}10^{-3}$ &
 $  1.6451{\times}10^{-3}$ &
 $  1.6451{\times}10^{-3}$ &
 $  5.0{\times}10^{-5}$ &
 $  5.0{\times}10^{-5}$ \\
  $B_{c}^{+}$ ${\to}$ $B_{s}^{{\ast}0}{K}^{+}$ & case 1b &
 $  2.9772{\times}10^{-3}$ &
 $  2.9766{\times}10^{-3}$ &
 $  2.9766{\times}10^{-3}$ &
 $ -1.9{\times}10^{-4}$ &
 $ -1.9{\times}10^{-4}$ \\
 $B_{c}^{+}$ ${\to}$ $B_{d}^{0}{\pi}^{+}$ & case 1b &
 $  3.7283{\times}10^{-3}$ &
 $  3.7272{\times}10^{-3}$ &
 $  3.7272{\times}10^{-3}$ &
 $ -3.0{\times}10^{-4}$ &
 $ -3.0{\times}10^{-4}$ \\
  $B_{c}^{+}$ ${\to}$ $B_{d}^{0}{\rho}^{+}$ & case 1b &
 $  5.2745{\times}10^{-3}$ &
 $  5.2742{\times}10^{-3}$ &
 $  5.2742{\times}10^{-3}$ &
 $ -5.0{\times}10^{-5}$ &
 $ -5.0{\times}10^{-5}$ \\
  $B_{c}^{+}$ ${\to}$ $B_{d}^{{\ast}0}{\pi}^{+}$ & case 1b &
 $  3.2682{\times}10^{-3}$ &
 $  3.2688{\times}10^{-3}$ &
 $  3.2688{\times}10^{-3}$ &
 $  1.9{\times}10^{-4}$ &
 $  1.9{\times}10^{-4}$ \\
  $B_{c}^{+}$ ${\to}$ $B_{d}^{0}{K}^{+}$ & case 1c &
 $  2.6616{\times}10^{-4}$ &
  --- & --- & --- & --- \\
  $B_{c}^{+}$ ${\to}$ $B_{d}^{0}{K}^{{\ast}+}$ & case 1c &
 $  2.2583{\times}10^{-4}$ &
  --- & --- & --- & --- \\
  $B_{c}^{+}$ ${\to}$ $B_{d}^{{\ast}0}{K}^{+}$ & case 1c &
 $  2.2075{\times}10^{-4}$ &
  --- & --- & --- & --- \\
  $B_{c}^{+}$ ${\to}$ $B_{u}^{+}{\overline{K}}^{0}$ & case 2a &
 $  2.2067{\times}10^{-5}$ &
  --- & --- & --- & --- \\
  $B_{c}^{+}$ ${\to}$ $B_{u}^{+}{\overline{K}}^{{\ast}0}$ & case 2a &
 $  1.8434{\times}10^{-5}$ &
  --- & --- & --- & --- \\
  $B_{c}^{+}$ ${\to}$ $B_{u}^{{\ast}+}{\overline{K}}^{0}$ & case 2a &
 $  1.8261{\times}10^{-5}$ &
  --- & --- & --- & --- \\
  $B_{c}^{+}$ ${\to}$ $B_{u}^{+}{\eta}$ & case 2b &
 $  1.5991{\times}10^{-6}$ &
 $  1.6122{\times}10^{-6}$ &
 $  1.6125{\times}10^{-6}$ &
 $  8.2{\times}10^{-3}$ &
 $  8.4{\times}10^{-3}$ \\
  $B_{c}^{+}$ ${\to}$ $B_{u}^{{\ast}+}{\eta}$ & case 2b &
 $  1.3042{\times}10^{-6}$ &
 $  1.2960{\times}10^{-6}$ &
 $  1.2964{\times}10^{-6}$ &
 $ -6.3{\times}10^{-3}$ &
 $ -6.0{\times}10^{-3}$ \\
  $B_{c}^{+}$ ${\to}$ $B_{u}^{+}{\pi}^{0}$ & case 2b &
 $  4.5968{\times}10^{-7}$ &
 $  4.5161{\times}10^{-7}$ &
 $  4.5134{\times}10^{-7}$ &
 $ -1.8{\times}10^{-2}$ &
 $ -1.8{\times}10^{-2}$ \\
  $B_{c}^{+}$ ${\to}$ $B_{u}^{+}{\rho}^{0}$ & case 2b &
 $  6.5030{\times}10^{-7}$ &
 $  6.4823{\times}10^{-7}$ &
 $  6.4776{\times}10^{-7}$ &
 $ -3.2{\times}10^{-3}$ &
 $ -3.9{\times}10^{-3}$ \\
 $B_{c}^{+}$ ${\to}$ $B_{u}^{+}{\omega}$ & case 2b &
 $  5.7921{\times}10^{-7}$ &
 $  5.8199{\times}10^{-7}$ &
 $  5.8212{\times}10^{-7}$ &
 $  4.8{\times}10^{-3}$ &
 $  5.0{\times}10^{-3}$ \\
  $B_{c}^{+}$ ${\to}$ $B_{u}^{{\ast}+}{\pi}^{0}$ & case 2b &
 $  4.0262{\times}10^{-7}$ &
 $  4.0722{\times}10^{-7}$ &
 $  4.0685{\times}10^{-7}$ &
 $  1.1{\times}10^{-2}$ &
 $  1.0{\times}10^{-2}$ \\
  $B_{c}^{+}$ ${\to}$ $B_{u}^{+}{\eta}^{\prime}$ & case 2d &
 $  8.8676{\times}10^{-8}$ &
 $  8.7700{\times}10^{-8}$ &
 $  8.7738{\times}10^{-8}$ &
 $ -1.1{\times}10^{-2}$ &
 $ -1.1{\times}10^{-2}$ \\
  $B_{c}^{+}$ ${\to}$ $B_{u}^{{\ast}+}{\eta}^{\prime}$ & case 2d &
 $  1.7401{\times}10^{-8}$ &
 $  1.7728{\times}10^{-8}$ &
 $  1.7731{\times}10^{-8}$ &
 $  1.9{\times}10^{-2}$ &
 $  1.9{\times}10^{-2}$ \\
  $B_{c}^{+}$ ${\to}$ $B_{u}^{+}{K}^{0}$ & case 2c &
 $  6.5428{\times}10^{-8}$ &
  --- & --- & --- & --- \\
  $B_{c}^{+}$ ${\to}$ $B_{u}^{+}{K}^{{\ast}0}$ & case 2c &
 $  5.4658{\times}10^{-8}$ &
  --- & --- & --- & --- \\
  $B_{c}^{+}$ ${\to}$ $B_{u}^{{\ast}+}{K}^{0}$ & case 2c &
 $  5.4143{\times}10^{-8}$ &
  --- & --- & --- & --- \\
 \end{tabular}
 \end{ruledtabular}
 \end{center}
 \end{table}


\begin{thebibliography}{99}
 \bibitem{cdf98} F. Abe, {\em et al.} (CDF Collaboration),
         Phys. Rev. {\bf D58}, 112004, (1998); 
         Phys. Rev. Lett. {\bf 81}, 2432, (1998). 
 \bibitem{cdf07} T. Aaltonen {\em et al.} (CDF Collaboration),
         arXiv:0712.1506 [hep-ex];
         A. Abulencia {\em et al.} (CDF Collaboration),
         Phys. Rev. Lett. {\bf 97}, 012002 (2006);
         Phys. Rev. Lett. {\bf 96}, 082002 (2006).
 \bibitem{d008} V. M. Abazov {\em et al.} (D0 Collaboration), arXiv:0802.4258 [hep-ex].
 \bibitem{0412158} N. Brambilla, {\em et al.} (Quarkonium Working Group),
          CERN-2005-005, hep-ph/0412158;
          M. P. Altarelli, F. Teubert, arXiv:0802.1901 [hep-ph].
 \bibitem{pan67p1559} I. P. Gouz, V. V. Kiselev, A. K. Likhoded, V. I. Romanovsky,
          O. P. Yushchenko, Phys. Atom. Nucl. {\bf 67}, 1559 (2004).
 \bibitem{prd71p074012} C. H. Chang, C. F. Qiao, J. X. Wang, X. G. Wu,
          Phys. Rev. {\bf D71}, 074012 (2005);
          Phys. Rev. {\bf D72}, 114009 (2005);
          C. H. chang, J. X. Wang, X. G. Wu, Phys. Rev. {\bf D77}, 014022 (2008);
          V. A. Saleev, D. V. Vasin, Phys. Lett. {\bf B605} 311, (2005);
          A. K. Likhoded, V. A. Saleev, D. V. Vasin,
          Phys. Atom. Nucl. {\bf 69} 94, (2006).
 \bibitem{lhc} http://ab-div.web.cern.ch/ab-div/Publications/LHC-DesignReport.html
 \bibitem{pdg2006} W. N. Yao {\em et al.}, J. Phys. {\bf G33}, 1 (2006).
 \bibitem{ckm} N. Cabibbo, Phys. Rev. Lett. {\bf 10}, 531, (1963);
          M. Kobayashi, and T. Maskawa, Prog. Theor. Phys. {\bf 49}, 652, (1973).
 \bibitem{ijmpa21p777} C. H. Chang, Int. J. Mod. Phys. {\bf A21}, 777 (2006).
 \bibitem{07070919} X. Liu, X. Q. Li, Phys. Rev. {\bf D77}, 096010 (2008). 
 \bibitem{prd75p097304} A. K. Giri, B. Mawlong, R. Mohanta,
          Phys. Rev. {\bf D75}, 097304 (2007); 
          Erratum  ibid. {\bf D76}, 099902 (2007);
          A. K. Giri, R. Mohanta, M. P. Khanna,
          Phys. Rev. {\bf D65}, 034016 (2002);
          V. V. Kiselev, J. Phys. {\bf G30}, 1445 (2004). 
 \bibitem{prd73p054024} M. A. Ivanov, J. G. Korner, P. Santorelli,
          Phys. Rev. {\bf D73}, 054024 (2006). 
 \bibitem{epjc45p711} J. F. Cheng, D. S. Du, C. D. L\"{u},
          Eur. Phys. J. {\bf C45}, 711 (2006). 
 \bibitem{prd70p074022} S. Fajfer, J. F. Kamenik, P. Singer,
          Phys. Rev. {\bf D70}, 074022 (2004). 
 \bibitem{prd68p094020} E. Ebert, R. N. Faustov, V. O. Galkin,
          Phys. Rev. {\bf D68}, 094020 (2003). 
 \bibitem{epjc32p29} E. Ebert, R. N. Faustov, V. O. Galkin,
          Eur. Phys. J. {\bf C32}, 29 (2003). 
 \bibitem{jpg28p595} V. V. Kiselev, O. N. Pakhomova, V. A. Saleev,
          J. Phys. {\bf G28},595 (2002). 
 \bibitem{jpg28p2241} G. L. Castro, H. B. Mayorga, J. H. Munoz,
          J. Phys. {\bf G28}, 2241 (2002) 
 \bibitem{prd65p114007} R. C. Verma, A. Sharma,
          Phys. Rev. {\bf D65}, 114007 (2002);
          Phys. Rev. {\bf D64}, 114018 (2001).
 \bibitem{pan64p1860} V. V. Kiselev, hep-ph/0211021;
          V. V. Kiselev, A. E. Kovalsky, A. K. Likhoded,
          Phys. Atom. Nucl. {\bf 64}, 1860 (2001);
          Nucl. Phys. {\bf B585}, 353 (2000). 
 \bibitem{pan64p2027} V. A. Saleev, Phys. Atom. Nucl. {\bf 64}, 2027 (2001); 
          O. N. Pakhomova, V. A. Saleev, Phys. Atom. Nucl. {\bf 63}, 1999 (2000).
 \bibitem{prd61p034012} P. Colangelo, F. D. Fazio,
          Phys. Rev. {\bf D61}, 034012 (2000). 
 \bibitem{prd62p057503} R. Fleischer, D. Wyler,
          Phys. Rev. {\bf D62}, 057503 (2000). 
 \bibitem{prd62p014019} A. A. El-Hady, J. H. Munoz, J. P. Vary,
          Phys. Rev. {\bf D62}, 014019 (2000). 
 \bibitem{cpl18p498} L. B. Guo, D. S. Du, Chin. Phys. Lett. {\bf 18}, 498 (2001).
 \bibitem{epjc9p557} Y. S. Dai, D. S. Du, Eur. Phys. J. {\bf C9}, 557 (1999).
 \bibitem{epjc5p705} D. S. Du, Z. T. Wei, Eur. Phys. J. {\bf C5}, 705 (1998).
 \bibitem{prd56p4133} J. F. Liu, K. T. Chao, Phys. Rev. {\bf D56}, 4133 (1997).
 \bibitem{plb387p187} D. S. Du, G. R. Lu, Y. D. Yang, Phys. Lett. {\bf B387}, 187 (1996).
 \bibitem{pan60p1729} A. V. Berezhnoi, V. V. Kiselev, A. K. Likhoded, A. I. Onishchenko,
          Phys. Atom. Nucl. {\bf 60}, 1729 (1997); 
          S. S. Gershtein, V. V. Kiselev, A. K. Likhoded, A. V. Tkabladze,
          A. V. Berezhnoi, A. I. Onishchenko, hep-ph/9803433.
 \bibitem{9605451} V. V. Kiselev, hep-ph/9605451.
 \bibitem{9504319} S. S. Gershtein, V. V. Kiselev, A. K. Likhoded, A. V. Tkabladze,
         Phys. Usp. {\bf 38}, 1 (1995). [hep-ph/9504319]
 \bibitem{prd49p3399} C. H. Chang, Y. Q. Chen, Phys. Rev. {\bf D49}, 3399 (1994).
 \bibitem{prd46p3836} Q. P. Xu, A. N. Kamal, Phys. Rev. {\bf D46}, 3836 (1992).
 \bibitem{plb286p160} M. Masetti, Phys. Lett. {\bf B286}, 160 (1992).
 \bibitem{prd39p1342} D. S. Du, Z. Wang, Phys. Rev. {\bf D39}, 1342, (1989).
 \bibitem{rmp68p1125} For a review, see G. Buchalla, A. J. Buras, M. E. Lautenbacher,
         Rev. Mod. Phys. {\bf 68}, 1125, (1996); 
         or A. J. Buras, hep-ph/9806471.
 \bibitem{bsw} M. Wirbel, B. Stech, and M. Bauer,
         Z. Phys. {\bf C29}, 637, (1985);
         M. Bauer, B. Stech, and M. Wirbel,
         Z. Phys. {\bf C34}, 103, (1987).
 \bibitem{9905312} M. Beneke, G. Buchalla, M. Neubert and C. T. Sachrajda,
          Phys. Rev. Lett. {\bf 83}, 1914, (1999); 
          Nucl. Phys. {\bf B591}, 313, (2000). 
 \bibitem{0108141} D. S. Du, H. J. Gong, J. F. Sun, D. S. Yang, and G. H. Zhu,
          Phys. Rev. {\bf D65}, 074001, (2002); 
          Phys. Rev. {\bf D65}, 094025, (2002); 
          and Erratum, ibid. {\bf D66}, 079904, (2002);
 \bibitem{prd68p054003} J. F. Sun, G. H. Zhu, D. S. Du,
          Phys. Rev. {\bf D68}, 054003, (2003).
 \bibitem{npb657p333} M. Beneke, M. Neubert,
          Nucl. Phys. {\bf B675}, 333, (2003).
 \bibitem{prd58p114006} Th. Feldmann, P. Kroll, B. Stech,
          Phys. Rev. {\bf D58}, 114006, (1998); 
 \bibitem{prd58p094009} A. Ali, G. Kramer, and C. D. L\"{u},
          Phys. Rev. {\bf D58}, 094009, (1998). 
 \bibitem{zpc57p43} P. Colangelo, G. Nardulli, N. Paver,
          Z. Phys. {\bf C57}, 43, (1993).
 \bibitem{9810339} D. Choudhury, A. Kundu, B. Mukhhopadhyaya, hep-ph/9810339.
 \bibitem{jpg26p1079} M. A Nobes, R M Woloshyn, J. Phys. {\bf G26}, 1079 (2000).
 \bibitem{prd63p074010} M. A. Ivanov, J. G. K\"{o}mer, P. Santorelli,
          Phys. Rev. {\bf D63}, 074010, (2001).
 \bibitem{mpla16p1439} D. Choudhury, A. Kundu, B. Mukhhopadhyaya,
          Mod. Phys. Lett. {\bf A16}, 1439 (2001).
 \bibitem{epjc47p413} T. M. Aliev, M. Savci, Eur. Phys. J. {\bf C47}, 413 (2006).
 \bibitem{epjc51p833} T. Huang, F. Zuo, Eur. Phys. J. {\bf C51}, 833 (2007).
 \bibitem{prd71p014029} P. Ball, R. Zwicky, Phys. Rev. {\bf D71}, 014029 (2005).
 \end{thebibliography}
 \end{document}